%% file: main.tex
\documentclass[sigconf]{acmart}

\usepackage{amsfonts}
\usepackage{algorithm}
\usepackage{algorithmic}
\usepackage{graphicx}
\usepackage{xcolor}

\usepackage{xspace}
\usepackage{color}

\usepackage{amssymb}
\usepackage{enumitem}
\usepackage{multirow}
\usepackage{subfigure}
\usepackage{threeparttable}

\usepackage{amsthm}

\newtheorem{thm}{Theorem}

\usepackage{subcaption}
\newcommand{\ours}[0]{FedCIA\xspace}

\AtBeginDocument{%
  }

\copyrightyear{2025}
\acmYear{2025}
\setcopyright{acmlicensed}
\acmConference[SIGIR '25]{Proceedings of the 48th International ACM SIGIR Conference on Research and Development in Information Retrieval}{July 13--18, 2025}{Padua, Italy}
\acmBooktitle{Proceedings of the 48th International ACM SIGIR Conference on Research and Development in Information Retrieval (SIGIR '25), July 13--18, 2025, Padua, Italy}
\acmDOI{10.1145/3726302.3729977}
\acmISBN{979-8-4007-1592-1/2025/07}

\begin{document}

\title{FedCIA: Federated Collaborative Information Aggregation for Privacy-Preserving Recommendation}

\author{Mingzhe Han}
\orcid{0000-0002-4911-6093}
\affiliation{
  \institution{Fudan University}
  \city{Shanghai}
  \country{China}
}
\email{mzhan22@m.fudan.edu.cn}

\author{Dongsheng Li}
\orcid{0000-0003-3103-8442}
\affiliation{
  \institution{Microsoft Research Asia}
  \city{Shanghai}
  \country{China}
}
\email{dongshengli@fudan.edu.cn}

\author{Jiafeng Xia}
\orcid{0000-0002-3777-3395}
\affiliation{
  \institution{Fudan University}
  \city{Shanghai}
  \country{China}
}
\email{jfxia19@fudan.edu.cn}

\author{Jiahao Liu}
\orcid{0000-0002-5654-5902}
\affiliation{%
  \institution{Fudan University}
  \city{Shanghai}
  \country{China}
}
\email{jiahaoliu21@m.fudan.edu.cn}

\author{Hansu Gu}
\orcid{0000-0002-1426-3210}
\affiliation{
  \city{Seattle}
  \country{United States}
}
\email{hansug@acm.org}

\author{Peng Zhang}
\orcid{0000-0002-9109-4625}
\authornote{Corresponding author.}
\affiliation{
  \institution{Fudan University}
  \city{Shanghai}
  \country{China}
}
\email{zhangpeng\_@fudan.edu.cn}

\author{Ning Gu}
\orcid{0000-0002-2915-974X}
\affiliation{
  \institution{Fudan University}
  \city{Shanghai}
  \country{China}
}
\email{ninggu@fudan.edu.cn}

\author{Tun Lu}
\orcid{0000-0002-6633-4826}
\authornotemark[1]
\affiliation{
  \institution{Fudan University}
  \city{Shanghai}
  \country{China}
}
\email{lutun@fudan.edu.cn}

\renewcommand{\shortauthors}{Mingzhe Han et al.}

\input{0-abstract}





\maketitle

\input{1-Introduction}
\input{2-Related-Work}
\input{2.5-Preliminaries}
\input{3-Method}
\input{4-Experiment}
\input{5-Conclusion}
\input{Acknowledgments}

\bibliographystyle{ACM-Reference-Format}
\bibliography{ref}

\end{document}

%% file: 0-Abstract.tex
\begin{abstract}
Recommendation algorithms rely on user historical interactions to deliver personalized suggestions, which raises significant privacy concerns. Federated recommendation algorithms tackle this issue by combining local model training with server-side model aggregation, where most existing algorithms use a uniform weighted summation to aggregate item embeddings from different client models. This approach has three major limitations: 1) information loss during aggregation, 2) failure to retain personalized local features, and 3) incompatibility with parameter-free recommendation algorithms.
To address these limitations, we first review the development of recommendation algorithms and recognize that their core function is to share collaborative information, specifically the global relationship between users and items. With this understanding, we propose a novel aggregation paradigm named collaborative information aggregation, which focuses on sharing collaborative information rather than item parameters.
Based on this new paradigm, we introduce the federated collaborative information aggregation (\ours) method for privacy-preserving recommendation. This method requires each client to upload item similarity matrices for aggregation, which allows clients to align their local models without constraining embeddings to a unified vector space. As a result, it mitigates information loss caused by direct summation, preserves the personalized embedding distributions of individual clients, and supports the aggregation of parameter-free models. 
Theoretical analysis and experimental results on real-world datasets demonstrate the superior performance of \ours compared with the state-of-the-art federated recommendation algorithms.
Code is available at https://github.com/Mingzhe-Han/FedCIA.
\end{abstract}

%% file: 1-Introduction.tex
\section{Introduction}
\label{sec:intro}

Recommendation algorithms~\cite{covington2016deep, liu2025enhancing, liu2024filtering, liu2025enhancing1,li2024recommender} aim to recommend items that users may be interested in based on their historical interactions. However, these interactions are private and regulations such as GDPR~\cite{regulation2018general} have been established to protect them, making it challenging for data holders to share information for unified model training. Federated recommendation algorithm~\cite{sun2024survey, tan2020federated, yang2020federated} addresses this issue by treating each data holder (e.g., users or companies) as a client. It allows these clients to train models on their own devices and then upload the trained models to a central server for aggregation. This approach enables the development of a reliable recommendation model while protecting user privacy.


Recent federated recommendation algorithms~\cite{perifanis2022federated, zhang2023dual, zhang2024gpfedrec} typically follow a unified aggregation paradigm. Each client initially trains their recommendation model based on their data. To protect user privacy, the embedding network is divided into user embeddings and item embeddings, where user embeddings are kept locally and item embeddings are uploaded and aggregated on a central server by specific algorithms such as weighted average~\cite{mcmahan2017communication}. The aggregated item embeddings are then distributed back to each client for the next round of training. We refer to this aggregation paradigm as \textbf{Weighted Summation Aggregation}, which aligns item embeddings across different clients into a common vector space, enabling clients to leverage shared information effectively. 

\begin{figure*}[t]
\centering
\includegraphics[width=2\columnwidth]{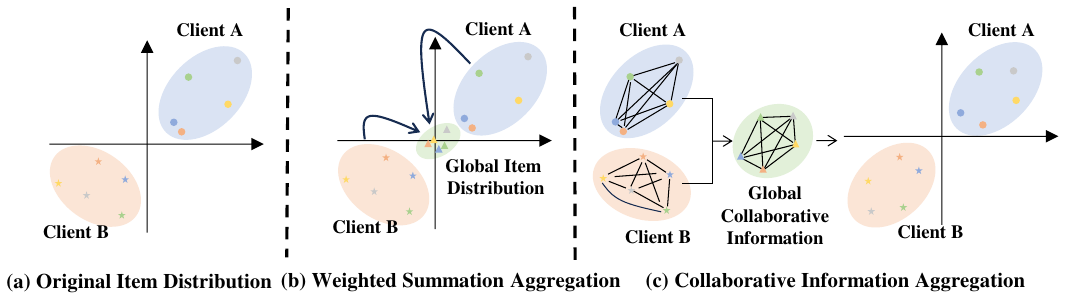}
\caption{
Illustration of two aggregation paradigms in federated recommendation algorithms.
}
\label{fig:agg}
\end{figure*}


Although this aggregation algorithm has been widely successful in various federated recommendation algorithms, it presents three issues.
\textbf{Firstly, the summation of item embeddings can lead to information loss.} For example, as illustrated in Figure~\ref{fig:agg}, two clients, A and B, train their recommendation algorithm models with their own datasets, resulting in opposite item vector spaces. Directly summing these vectors will result in the aggregated server embeddings that tend to be zero-vectors, which lack useful information.
\textbf{Secondly, there is a lack of personalized user modeling.} After client aggregation, the server distributes unified items to each client, ensuring all client items are in a consistent vector space. However, this global distribution space is not optimal for every client, it will result in a loss of personalized modeling for specific users, leading to suboptimal results. Some personalized federated recommendation systems~\cite{zhang2023dual, zhang2024gpfedrec}, achieve personalization through dual model and optimization. However, they still rely on item embeddings for aggregation, which can lead to reduced convergence efficiency and effectiveness due to information loss during aggregation.
\textbf{Finally, they are unable to accommodate parameter-free Collaborative Filtering (CF) models, e.g. Graph Signal Processing.} These algorithms require that all clients use the parameter-based model. However, in real-world scenarios, the diversity of devices and user requirements often demands varied model structures, necessitating federated learning frameworks to have greater scalability.





In this paper, we suggest that the root cause of these challenges lies in the over-reliance on the model parameter during the aggregation.
To address this issue, we revisit the evolution of recommendation algorithms over the years and observe a critical oversight. 
The superior performance of deep learning~\cite{he2017neural, he2020lightgcn, xia2022hypergraph, liu2023autoseqrec} has encouraged researchers to parameterize users and items, while ignoring the collaborative information that played a fundamental role in early recommendation algorithms~\cite{herlocker1999algorithmic, hu2008collaborative}. Here the collaborative information indicates the relationship between all users and items.
Therefore, when parameter aggregation within the federated learning framework results in information loss, it becomes essential to reconsider the importance of collaborative information.
To this end, we propose a novel federated learning aggregation paradigm named \textbf{Collaborative Information Aggregation}. As illustrated in Figure~\ref{fig:agg}, this paradigm no longer requires the summation of model parameters in federated learning. Instead, it focuses on sharing collaborative information (e.g. item similarity~\cite{sarwar2001item}) as modeled by their respective recommendation algorithms.


Based on this aggregation paradigm, we propose a novel federated recommendation algorithm framework named Federated Collaborative Information Aggregation (\ours). In this framework, we choose item similarity as the collaborative information in aggregation, thus each client uploads item similarity matrices instead of item embeddings. The algorithm aggregates these similarity matrices to derive a global item similarity matrix, which is then distributed to each client. Each client optimizes its local model to align its local item distribution with the global item.
On one hand, as shown in Figure~\ref{fig:agg}, this framework prevents information loss due to varying item distributions during aggregation. On the other hand, this framework avoids forcing item embeddings into a uniform vector space, thus preserving the personalized information of each client. In addition, this aggregation does not require parameter uploads, making it ideal for aggregation among parameter-free models.
We also provide a theoretical analysis from the perspective of graph signal processing, demonstrating that aggregating item similarity optimally captures collaborative information.

Our main contributions are summarized as follows:
\begin{itemize}
\item We introduce a novel aggregation framework \ours for federated recommendation algorithms. To our knowledge, this is the first federated learning work that aggregates collaborative information instead of model parameters.
\item \ours does not need to upload model parameters, which can be utilized in parameter-free collaborative filtering algorithms. To our knowledge, this is the first federated learning work compatible with parameter-free recommendation algorithms.
\item We propose to leverage the sum of similarity matrices to aggregate collaborative information and provide a theoretical analysis to support this algorithm.
\item Experimental results on real-world datasets demonstrate that \ours achieves higher accuracy compared to existing state-of-the-art federated recommendation algorithms.
\end{itemize}

%% file: 2-Related-Work.tex
\section{Related Work}

\subsection{Federated Learning}

Federated Learning (FL) \cite{yang2019federated, li2020review, li2020federated} is a distributed framework enabling multiple clients to collaboratively train a central model while retaining their raw data locally.
The foundational algorithm, FedAvg \cite{mcmahan2017communication}, aggregates client models by weighted summing them based on the number of samples in each dataset.
Subsequent research has primarily focused on server-side aggregation and client-side training. For instance, FedAvgm \cite{hsu2019measuring} introduces a momentum module that incorporates previous aggregation rounds to enhance model updates. FedProx \cite{li2020heterogeneous} constrains parameter changes during each client's training round by using a loss function.
Furthermore, given the practical deployment of federated learning, some studies have explored heterogeneous model aggregation \cite{li2019fedmd, lu2021heterogeneous, ye2023heterogeneous}. 
These studies have not specifically addressed recommendation algorithms, where data distribution is more dispersed and certain user-related parameters pose privacy concerns.

\subsection{Federated recommendation algorithms}

The federated recommendation algorithm~\cite{sun2024survey, tan2020federated, yang2020federated, perifanis2022federated, zhang2023dual, zhang2024gpfedrec} integrates the federated learning framework into recommendation models while incorporating specific designs tailored to the unique requirements of these systems.
FedRec~\cite{lin2020fedrec} mitigates privacy risks by randomly sampling unrated items and assigning virtual ratings. This approach prevents servers from inferring user interactions based on gradient information.
FedMF~\cite{chai2020secure} employs homomorphic encryption during the aggregation process, ensuring that user ratings remain private while allowing the server to aggregate gradients securely. 
FedNCF~\cite{perifanis2022federated} extended this work by applying the commonly used Neural Collaborative Filtering (NCF) framework, partitioning parameters into user-related and item-related components. Only the item-related parameters are uploaded during aggregation. 
While these approaches successfully achieve model aggregation while safeguarding user privacy, they impose a uniform parameter structure across all clients, disregarding individual user preferences. This limitation often leads to suboptimal solutions by failing to account for personalized information.

Recognizing the importance of personalization in federated recommendation algorithms, recent research has explored personalized federated learning approaches. PFedRec~\cite{zhang2023dual} introduces a dual personalization mechanism that optimizes both the global model and individual client models to provide user-specific recommendations. GPFedRec~\cite{zhang2024gpfedrec} takes a different approach by constructing a user relationship graph to capture inter-user correlations and leveraging local embeddings to model personalized features. While these methods address user personalization, they still rely on weighted summation of item features during aggregation. This aggregation strategy can lead to the loss of collaborative information, ultimately reducing the quality of the final model. In contrast, our proposed method effectively mitigates this issue by preserving collaborative information during aggregation, ensuring more robust and accurate recommendations.

%% file: 2.5-Preliminaries.tex
\section{Preliminaries}
\label{sec:pre}

\subsection{Recommendation Algorithms}

Recommendation algorithms aim to predict the user-item interaction $y$ between a user $u$ and an item $i$ based on the user history dataset $\mathcal{D} = (\mathcal{U}, \mathcal{I}, \mathcal{Y})$, where $\mathcal{U}$ represents the set of users, $\mathcal{I}$ represents the set of items, and $\mathcal{Y}$ represents the interaction labels. The label $y$ is a binary variable ($y \in \{0, 1\}$) indicating whether the user $u$ is interested in the item $i$ ($y=1$) or not ($y=0$).

Most recommendation algorithms $R(.)$ follow the same paradigm. First, the user $u$ and the item $i$ are mapped into embeddings through embedding networks $E_u(.|\theta_u)$ and $E_i(.|\theta_i)$, respectively. Then the embeddings are used to compute a prediction score $\hat{y}$ through a score network $S(.|\theta_s)$. Formally, the recommendation algorithm can be expressed as $\hat{y} = R(u,i) = S(E(u,i))$ where $E(u,i)$ represents the combined embeddings of the user and item, and $S(.)$ is the function used to generate the prediction score.
The model parameters $\theta$ of the recommendation algorithm $R(.)$ are divided into three components: the user embedding parameters $\theta_u$, the item embedding parameters $\theta_i$ and the score network parameters $\theta_s$.
In some recommendation algorithms, the prediction score is obtained directly from the dot product of the user and item embeddings. In such cases, the score network $\theta_s$ does not include any additional parameters, simplifying the model architecture.

\subsection{Federated Recommendation Algorithms}

In federated learning, users store their interaction history on their own devices to ensure privacy. We assume that there are $K$ user clusters, each containing one or more users. Thus, the dataset for federated learning consists of $K$ sets of data samples distributed across $K$ clients, denoted as $\mathcal{D}_k = (\mathcal{U}, \mathcal{I}, \mathcal{Y})_k$, where $k \in [1, K]$. The goal of federated learning is to train a global model $R(.)$ while protecting the interaction data of each user.
In conventional federated learning, such as FedAvg~\cite{mcmahan2017communication}, each client first trains its local model based on its local dataset $\mathcal{D}_k$, The model parameters are then uploaded to a server for aggregation using a weighted average algorithm, based on the number of samples in each dataset: $R(.|\theta) = \sum_{k=1}^{K} \frac{|\mathcal{D}_k|}{|\mathcal{D}|} R_k(.|\theta)$. 

In recommendation algorithms, due to the unique properties of parameters, a different aggregation algorithm is used. User embeddings $\theta_u$ are directly linked to individual users, and clients keep these embeddings local to maintain privacy. Instead, upload and aggregate other parameters $\theta / \theta_u$. This approach facilitates information sharing among clients while preserving user privacy, effectively training a global federated recommendation algorithm model.

\subsection{Graph Signal Processing}
Given a graph $\mathcal{G} = \{\mathcal{V}, \mathcal{E}\}$ comprising $N$ nodes, it can be represented by an adjacency matrix $A$, where $A_{ij}$ indicates the presence of a relationship between nodes $i$ and $j$. The normalized form of the adjacency matrix is denoted as $\tilde{A}=D^{1/2}AD^{-1/2}$, where $D$ is the degree matrix.
In the field of graph signal processing, filters are crucial tools used to uncover complex relationships between nodes and accurately predict missing links. Filters are primarily designed based on the (normalized) Laplacian matrix~\cite{chung1997spectral, spielman2007spectral} $\tilde{L}=I-\tilde{A}$, expressed as: $F=f(\tilde{L}){=}f(U\Sigma U^\top){=}U\text{Diag}([f(\lambda_1),\cdots,f(\lambda_N)])U^T$ where $f(.)$ represents the frequency response function, $U$ and $\Sigma$ denote the eigenvector matrix and eigenvalue matrix of $\tilde{L}$, respectively. Different frequency response functions can construct various filters, thereby uncovering different relationships between nodes.

Graph signal processing is widely applied in recommendation systems\cite{liu2022parameter, liu2023personalized, liu2023triple, liu2023recommendation, liu2025mitigating, xia2022fire, xia2024hierarchical, xia2024neural} for its significant effectiveness and efficiency.
In recommendation algorithms, users and items are represented as nodes in a bipartite graph, and their interactions are modeled as edges. Generally, the graph is formally represented using an adjacency matrix $A \in \{0,1\}^{N \times M}$, where $N$ is the number of users and $M$ is the number of items. Each entry $A_{ij} = 1$ indicates that user $i$ has interacted with item $j$; otherwise, $A_{ij} = 0$.

The filter $F$ in graph signal processing methods is typically an $M\times M$ matrix that encodes the relationships or correlations between items. The simplest filter that those methods adopt is the linear filter, which aggregates the interaction matrix after normalization. Specifically, the filter can be expressed as $F = \tilde{A}^T\tilde{A}$, where $\tilde{A}$ is the normalized interaction matrix. The normalization process is defined as $\tilde{A_s} = D_u^{-1/2}AD_i^{-1/2}$, where $D_u$ is the degree matrix of the users and $D_i$ is the degree matrix of the items. 

\section{Theoretical Analysis}
\label{sec:the}

In this section, we first propose that, for parameter-based models, the generated item similarity can be interpreted as a linear filter within the framework of graph signal processing. This enables the application of theoretical analysis from graph signal processing to both parameter-based and parameter-free collaborative filtering models.
We then emphasize that the average of item similarities can approximate global item similarity. Finally, we identify the limitations of the existing weighted summation aggregation.

\subsection{Item Similarity and Linear Filters}

In parameter-based recommendation algorithm methods, items are mapped to item embeddings, which encapsulate the model's representation of item features. Intuitively, the similarity between items, computed as the dot product of their corresponding item embeddings, can be interpreted as a linear filter.
Taking Matrix Factorization (MF) as an example, it decomposes the interaction matrix into user embeddings and item embeddings and optimizes these embeddings to find the optimal solution by minimizing a loss function. One of the optimal solutions can be understood through Singular Value Decomposition (SVD). Specifically, SVD decomposes the interaction matrix $A$ into $U \Sigma_{\tilde{A}} V^T $, where $\sqrt{\Sigma_{\tilde{A}}}V^T $ can be viewed as the item embeddings optimized by MF. The dot product of these item embeddings can then be expressed as $V^T\Sigma_{\tilde{A}} V$, which represents a linear filter whose frequency response function is $f(\lambda)=1-\lambda$ in this context.

\subsection{Collaborative Information Aggregation}

We begin by analyzing the aggregation of collaborative information, with a particular emphasis on the impact of privacy considerations in this process.
We analyzed centralized learning without considering privacy and federated learning with privacy considerations and proposed the following theorem.

\begin{thm}
In a federated recommendation algorithm, assuming that the popularity of each item (i.e., the number of user interactions) is identical, the average of each local linear filter is equivalent to the global ideal linear filter. 
\end{thm}

Note that the 'global ideal linear filter' here refers to a global linear filter that does not consider privacy risks, not an 'ideal filter'. And here is the proof.

\begin{proof}
Consider the complete global interaction matrix as $A_s\in\{0,1\}^{N \times M}$. In an ideal scenario where no privacy risks are considered, complete user data is processed on a central server, and the linear filter is:
\begin{equation}
\begin{aligned}
F_{ideal} &= \tilde{A_s}^T \tilde{A_s} = D_i^{-1/2}A_sD_u^{-1}A_sD_i^{-1/2}.
\end{aligned}
\end{equation}

In reality, due to privacy risks, data is stored on each client.
The interaction matrix is split among $N$ clients, where client $k$ holds $A_k \in \{0,1\}^{1 \times M}$.
Each client generates a local linear filter $F_k = \tilde{A_k^T}\tilde{A_k}$ using its interaction matrix. The average of these $N$ filters are:
\begin{equation}
\begin{aligned}
F_{agg} &= \frac{1}{N} (F_1 + F_2 + ... + F_N) \\
&= \frac{1}{N} (D_{i1}^{-1/2}A_1D_{u1}^{-1}A_1D_{i1}^{-1/2} + D_{i2}^{-1/2}A_2D_{u2}^{-1}A_2D_{i2}^{-1/2} + ... \\ & + D_{iN}^{-1/2}A_ND_{uN}^{-1}A_ND_{iN}^{-1/2}) \\ 
&=  \frac{1}{N} (A_1D_{u1}^{-1}A_1 + A_2D_{u2}^{-1}A_2 + ... + A_ND_{uN}^{-1}A_N) \\
&= \frac{1}{N} [A_1||A_2||...||AN]^T [D_{u1}||D_{u2}||...||D_{uN}][A_1||A_2||...||AN] \\
&= \frac{1}{N} A_sD_u^{-1}A_s.
\end{aligned}
\end{equation}

In federated learning, counting the exact number of interactions per item $D_i$ is challenging due to privacy concerns. Assuming each item is interacted with once, we set $D_i=NI$. Thus, the ideal linear filter becomes:
\begin{equation}
\begin{aligned}
F_s &= D_i^{-1/2}A_sD_u^{-1}A_sD_i^{-1/2} = \frac{1}{N} A_sD_u^{-1}A_s = F_{agg}.
\end{aligned}
\end{equation}

This shows that under the assumption of equal item popularity, the average of local linear filters equals the global ideal linear filter.

\end{proof}

Therefore, by considering item similarity as a filter in graph signal processing, the global filter can approximate the mean of multiple local filters. This approach enables the aggregation of item similarity matrices from individual clients to approximate a global item similarity matrix.

\begin{figure*}[t]
\centering
\includegraphics[width=2\columnwidth]{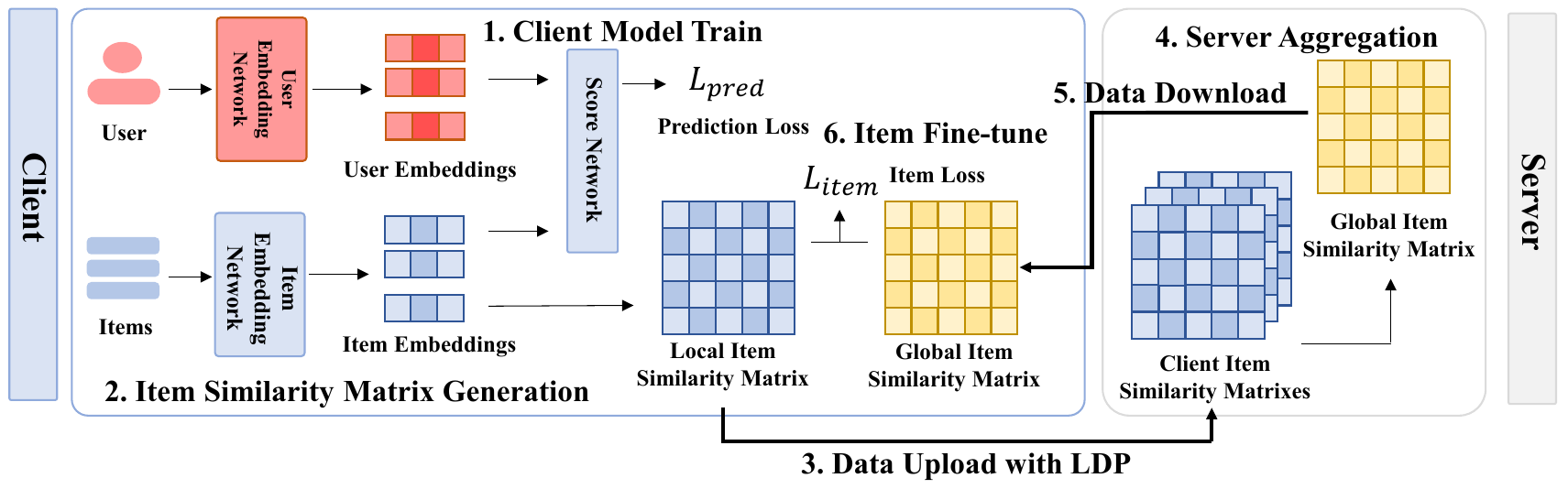}
\caption{
The illustration of the \ours framework. We only illustrate a single client for simplicity. The red part indicates private information and the yellow part indicates global information.
}
\label{fig:pb}
\end{figure*}

\subsection{Loss Information during Aggregation}

In the previous section, we demonstrated that item similarity serves as an effective medium for information aggregation. In this section, we analyze the impact of different aggregation paradigms on global information. For simplicity and consistency, we adopt the L1 distance as the similarity metric between items, as it is used in clustering~\cite{kashima2008k, ye2017l1} and metric learning~\cite{wang2014robust}.
The L1 distance, where larger values indicate weaker correlations, is used here to evaluate the amount of information captured by the correlation matrix. Since the L1 distance is always a positive scalar, we posit that a correlation matrix with a broader range of values conveys more information.

Consider two clients, A and B, each holding the embeddings of two items. Let the embeddings be $e_{a1}, e_{a2}$ for client A and $e_{b1}, e_{b2}$ for client B. 
In weighted summation aggregation, the global embeddings are obtained by directly summing the embeddings from both clients. The recommendation algorithm then computes the similarity between items as $S_{WS} = \frac{1}{2}|(e_{a1} + e_{b1}) - (e_{a2} + e_{b2})|.$

In contrast, collaborative information aggregation first computes item similarities locally before aggregation. By the triangle inequality, we have:
\begin{equation}
\begin{aligned}
S_{CI} &= \frac{1}{2}|e_{a1} - e_{a2}| + \frac{1}{2}|e_{b1} - e_{b2}| \\ 
&\ge \frac{1}{2}|e_{a1} - e_{a2} + e_{b1} - e_{b2}| \\
&= \frac{1}{2}|(e_{a1} + e_{b1}) - (e_{a2} + e_{b2})| = S_{WS}.
\end{aligned}
\end{equation}

This inequality highlights that weighted summation aggregation can result in information loss, particularly when the embeddings from clients A and B have opposite values. In such cases, weighted summation aggregation fails to retain meaningful information. By contrast, collaborative information aggregation consistently preserves more information due to its larger value range, making it a more robust approach for constructing the global collaborative information.

%% file: 3-Method.tex
\section{Methodology}
\label{sec:method}

In this section, we introduce the proposed \ours framework, which is versatile and applicable to various recommendation algorithms, including both parameter-based and parameter-free collaborative filtering models. We first outline the algorithm for parameter-based models, followed by alternative solutions for parameter-free models. Finally, we discuss the challenges our method addresses in practical deployment.

\subsection{Overview}

We first introduce the process of our framework under parameter-based models. Figure \ref{fig:pb} illustrates the complete process of \ours during a round of federated learning communication. Each client initially trains its model using local data. As described in Section~\ref{sec:the}, we use item similarity as the collaborative information for aggregation, and these terms (i.e. ``collaborative information'' and ``item similarity'') will be used interchangeably in this section. The local model then generates a local item similarity matrix based on the trained item embeddings, which is uploaded to the server. The server aggregates these matrices to form a global item similarity matrix. Finally, the client fine-tunes its item embeddings based on this global matrix.

\subsection{Training Pipeline}

\subsubsection{Data Training}
In the beginning, each client trains a recommendation model locally using its own data, formulated as:
\begin{equation}
\begin{aligned}
\label{equ:loss}
L_{pred} = \frac{1}{|\mathcal{D}|}\sum_{(u,i,y) \in (\mathcal{U}, \mathcal{I}, \mathcal{Y})}L(\theta; R(u,i), y),
\end{aligned}
\end{equation}
where we use BPR loss as the loss function.

\subsubsection{Data Upload}
As described in Section~\ref{sec:pre}, the recommendation algorithm model maps discrete item IDs to corresponding item embeddings. Based on these embeddings, the client computes the complete item similarity matrix $C = E(i|\theta_i) ^ TE(i|\theta_i)$ using a dot product. Each client $k$ then uploads its local similarity matrix $C_k$ as collaborative information to the server for aggregation.

\subsubsection{Data Aggregation}
Upon receiving the item similarity matrices from all clients, the server aggregates them to generate the global similarity matrix. In Section~\ref{sec:the}, we demonstrate that the average of local linear filters approximates the global linear filters. Leveraging this theorem, \ours aggregates the similarity matrices by taking the average of those uploaded by each client. Thus, the global similarity matrix at the server is: $C_s =  \frac{1}{N} \sum_{k=1}^{K} C_k.$

\subsubsection{Data Download}
Our data downloads differ from weighted summation aggregation, where each local client can directly substitute its local item embeddings with global item embeddings from the server to incorporate aggregated information. However, the parameters of the local model typically do not include collaborative information, such as the item similarity matrix, which is not explicitly part of most recommendation algorithms. Although we obtain the global item similarity matrix through aggregation algorithms, this matrix cannot directly replace the local model's parameters. Instead, we must approximate the local model's collaborative information to align with the global collaborative information.

Intuitively, item similarity is derived from their embeddings, so the global item similarity can be treated as a label to fine-tune the local item embeddings, ensuring that the local item similarity matrix approximates the global item similarity matrix. We choose the Mean Squared Error (MSE) as the loss function for this fine-tuning process. The loss for the entire fine-tuning process is expressed as:
\begin{equation}
\begin{aligned}
\label{equ:itemloss}
L_{item} = L(\theta_i;C_s) = \text{MSE}(E(i|\theta_i)^TE(i|\theta_i), C_s).
\end{aligned}
\end{equation}

\subsubsection{Privacy Protection}
In weighted summation aggregation methods, the privacy risks associated with uploading item embeddings were addressed by employing a Local Differential Privacy (LDP) strategy, where zero-mean Laplacian noise was added directly to the item embeddings. 
In our algorithm, we calculate the similarity matrix by dot products based on the item embeddings. This similarity matrix transmits less information because attackers can only infer item similarities but cannot directly recover the original item embeddings, thereby reducing the risk of information leakage.
Nevertheless, we also apply zero-mean Laplacian noise to the similarity matrix before uploading it to mitigate the potential risk as $C_i = C_i + Laplacian(0,  \delta)$,
where $\delta$ denotes the intensity of the Laplacian noise, which is derived from the privacy budget and the sensitivity of the similarity matrix.

We train the recommendation algorithm model through multiple rounds of communication. 
From the client's perspective, it initially trains a local recommendation algorithm model using its own data. It then generates the item similarity matrix locally based on this model. Finally, the client uploads the similarity matrix, enhanced with local differential privacy, to the server. 
On the server side, it aggregates the received item similarity matrices into a global similarity matrix by average. The complete model training pipeline is detailed in Algorithm~\ref{alg:training}.

\begin{algorithm}[tb]
\caption{\ours Training Pipeline}
\label{alg:training}
\begin{flushleft}
\textbf{Input}: Training set $D_k$ for each client $k$. \\
\textbf{Parameters}: 
Maximum number of communication rounds $com\_num$, maximum number of client item training epoch $epoch_{i}$, maximum number of client training epoch $epoch_{c}$, and the number of clients $K$.
\end{flushleft}
\begin{algorithmic}[1]

\FOR{$t$ = $1,\ldots,com\_num$}

\STATE \textbf{Server aggregation:}
\STATE Generate the global item similarity matrix $C_s = \frac{1}{N} \sum_{k=1}^{K} C_k$.

\STATE \textbf{Local training:}

\FOR{$k$ = $1,\ldots,K$}
\FOR{$epoch$ = $1,\ldots,epoch_{c}$}
    \FOR{$(u,i,y)$ in $(\mathcal{U}, \mathcal{I}, \mathcal{Y})$}
        \STATE Train client model by $L_{pred}$.
    \ENDFOR
\ENDFOR
\STATE Generate the item similarity matrix $C_k$.
\STATE Add Laplacian noise on item similarity matrix $C_k$.
\STATE Upload the item similarity matrix to the server.
\STATE Wait for \textbf{Server aggregation}.
\STATE Downloads global similarity matrix $C_s$.
\FOR{$epoch$ = $1,\ldots,epoch_{i}$}
    \STATE Train client item embeddings by $L_{item}$.
\ENDFOR
\ENDFOR

\ENDFOR
\end{algorithmic}
\end{algorithm}

\subsection{Parameter-free Model Aggregation}

In this section, we extend our aggregation method to parameter-free collaborative filtering models. Existing weighted summation aggregation techniques rely on parameter calculations, making them unsuitable for parameter-free models like graph signal processing~\cite{shen2021powerful}. In contrast, \ours can perform aggregation without requiring model parameters, allowing it to be adapted for use with parameter-free models.

\begin{table}[t]
\centering
\caption{The statistics of our datasets.
}
\resizebox{0.8\columnwidth}{!}
{
\begin{tabular}{l|ccc}
\toprule
datasets & \# Users & \# Items & \# Interactions \\
\midrule
Ml-100k  & 943      & 1682     & 100000          \\
Ml-1m    & 6040     & 3706     & 1000209         \\
Beauty (Amazon Beauty)  & 22363    & 12101    & 198502          \\
LastFM   & 992      & 10000    & 517817          \\
BX (Book-Crossing)       & 18964    & 19998    & 482153          \\
\bottomrule
\end{tabular}
}
\label{tab:dataset}
\end{table}

\subsubsection{Collaborative Information Aggregation}

In graph signal processing, the original interaction matrix is processed using various filters, and recommendations are generated based on the score rankings of the filtered results. To facilitate aggregation, we directly select the filter as the collaborative information to be shared. For algorithms that require a combination of multiple filters (e.g., GF-CF), we use the final mixed filter as the collaborative information to upload. Once the filters are uploaded, the server aggregates them into a global filter using an average approach, similar to the aggregation process for parameter-based models.

\input{table_result}

\subsubsection{Collaborative Information Utilization}

In graph signal processing, global collaborative information corresponds to a global filter $F_{agg}$, which each client can use to generate the final predicted interaction matrix. However, similar to parameter-based models, relying solely on global information may lead to a loss of personalized information. Therefore, we incorporate both the global filter and the individual filter $F_{k}$ of each client $k$, as $\hat{A_k} = A_k  F_k + \beta A_k  F_{agg}$, where $\hat{A_i}$ refers to the prediction of interaction matrix $A_i$ and $\beta$ is a hyperparameter.

\subsection{Discussion}

In this section, we discuss the practical deployment of our \ours, focusing on model scalability and communication resources.

\subsubsection{Model Scalability}
\label{sec:scal}

In real-world federated recommendation systems, user devices often have varying computing capabilities, leading to models with different architectures. This diversity poses a challenge to the scalability of federated learning algorithms. Traditional parameter-based weighted summation aggregation requires consistent model architectures across clients, as differences in model size result in incompatible parameter vectors.
Our proposed \ours addresses this issue by using collaborative information for aggregation, relying only on the similarity between different items. This similarity is typically obtained through the dot product of item embeddings. Consequently, \ours is applicable to all parameter-based recommendation models that encode items into embeddings.
In our experiments, we evaluated the aggregation results under heterogeneous model conditions. The findings demonstrate that our method effectively aggregates information across different model architectures and sizes, exhibiting enhanced scalability.

\subsubsection{Communication Resources}

Our method inherently requires more communication resources. Rather than uploading model parameters, our approach involves uploading the item similarity matrix, sized at $R^{M \times M}$. In contrast, item embeddings are sized at $R^{M \times L_E}$, where $L_E$ denotes the embedding dimension. Consequently, uploading the item similarity matrix directly often demands significantly more communication resources. However, in practice, a single training session typically does not utilize all items, and the item similarity matrix usually contains substantial noise. 
To mitigate communication overhead, we propose using truncated singular value decomposition to decompose the item similarity matrix, retaining only the top-$L_k$ singular values and corresponding vectors. Given that the similarity matrix is symmetric, the left singular vectors are identical to the right singular vectors, so only half need to be transmitted. The server then reconstructs the filters for each client using these vectors and calculates the global similarity matrix.
This method reduces the size of the uploaded data from the original $R^{M \times M}$ to $R^{M \times L_k} + R^{L_k}$, here $L_k$, similar to $L_E$, is approximately one percent of the item number $M$.
This approach achieves a level of communication efficiency comparable to uploading model parameters while preserving most of the crucial information.

%% file: table_result.tex
\begin{table*}[t]
\centering
\caption{The overall comparison for all baseline methods in five datasets. The boldface indicates the best result and the underline indicates the secondary.}
\resizebox{\textwidth}{!}
{
\begin{tabular}{c|c|ccc|ccccc|ccc}
\toprule
\multirow{2}{*}{Dataset} & \multirow{2}{*}{Metric} & \multicolumn{3}{c|}{No Aggregation} & \multicolumn{5}{c|}{Weighted Summation Aggregation} & \multicolumn{3}{c}{Collaborative Information Aggregation (Ours)}                                  \\ \cline{3-13} 
                         &                         & MF            & LightGCN        & GF-CF         & FedMF   & FedNCF & PFedrec & GPFedrec & FedLightGCN & $\text{FedCIA}_\text{MF}$ & $\text{FedCIA}_\text{LightGCN}$ & $\text{FedCIA}_\text{GF-CF}$ \\ \hline
\multirow{3}{*}{Ml-100k}  & F1@10                   & 0.0644        & 0.1012          & 0.1040        & 0.2074  & 0.1321 & 0.1428  & 0.1413   & 0.1828      & \underline{0.2326}              & 0.1881                          & \textbf{0.2356}              \\
                         & MRR@10                  & 0.2557        & 0.3760          & 0.3635        & 0.5578  & 0.4350 & 0.4750  & 0.4685   & 0.5653      & \textbf{0.6008}           & 0.5601                          & \underline{0.5806}                 \\
                         & NDCG@10                 & 0.3307        & 0.4558          & 0.4515        & 0.6460  & 0.5244 & 0.5577  & 0.5555   & 0.6284      & \textbf{0.6766}           & 0.6298                          & \underline{0.6704}                 \\ \hline
\multirow{3}{*}{Ml-1m}    & F1@10                   & 0.0471        & 0.0951          & 0.1264        & 0.1106  & 0.0951 & 0.0989  & 0.0964   & 0.1690      & \underline{0.1814}              & 0.1687                          & \textbf{0.1845}              \\
                         & MRR@10                  & 0.1826        & 0.3043          & 0.3797        & 0.3367  & 0.3098 & 0.3387  & 0.3362   & 0.4470      & \underline{0.4705}              & 0.4527                          & \textbf{0.4815}              \\
                         & NDCG@10                 & 0.2490        & 0.3940          & 0.4667        & 0.4203  & 0.3941 & 0.4125  & 0.4072   & 0.5368      & \underline{0.5561}              & 0.5403                          & \textbf{0.5647}              \\ \hline
\multirow{3}{*}{Beauty}  & F1@10                   & 0.0068        & 0.0222          & 0.0248        & 0.0068  & 0.0061 & 0.0182  & 0.0173   & 0.0245      & 0.0080                    & \underline{0.0250}                    & \textbf{0.0358}              \\
                         & MRR@10                  & 0.0117        & 0.0355          & 0.0402        & 0.0123  & 0.0109 & 0.0349  & 0.0325   & 0.0397      & 0.0135                    & \underline{0.0404}                    & \textbf{0.0571}              \\
                         & NDCG@10                 & 0.0172        & 0.0511          & 0.0576        & 0.0184  & 0.0165 & 0.0506  & 0.0478   & 0.0571      & 0.0206                    & \underline{0.0586}                    & \textbf{0.0821}              \\ \hline
\multirow{3}{*}{LastFM}  & F1@10                   & 0.0140        & 0.0315          & 0.0388        & 0.0398  & 0.0444 & 0.0439  & 0.0475   & 0.0606      & 0.0578                    & \textbf{0.0761}                 & \underline{0.0756}                 \\
                         & MRR@10                  & 0.2094        & 0.3003          & 0.3597        & 0.3408  & 0.4014 & 0.3666  & 0.4168   & 0.4792      & 0.4563                    & \underline{0.5272}                    & \textbf{0.5466}              \\
                         & NDCG@10                 & 0.2691        & 0.3859          & 0.4503        & 0.4337  & 0.4866 & 0.4617  & 0.5007   & 0.5670      & 0.5385                    & \underline{0.6172}                    & \textbf{0.6235}              \\ \hline
\multirow{3}{*}{BX}      & F1@10                   & 0.0038        & 0.0050          & 0.0069        & 0.0102  & 0.0122 & 0.0137  & 0.0123   & 0.0107      & 0.0115                    & \underline{0.0140}                    & \textbf{0.0331}              \\
                         & MRR@10                  & 0.0092        & 0.0119          & 0.0140        & 0.0217  & 0.0251 & \underline{0.0296}  & 0.0274   & 0.0233      & 0.0242                    & 0.0290                    & \textbf{0.0625}              \\
                         & NDCG@10                 & 0.0136        & 0.0171          & 0.0202        & 0.0321  & 0.0374 & \underline{0.0426}  & 0.0393   & 0.0337      & 0.0358                    & 0.0422                    & \textbf{0.0885}             \\
\bottomrule
\end{tabular}
}
\label{tab:result}
\end{table*}

%% file: 4-Experiment.tex
\section{Experiments}

In this section, we introduce and analyze the following research questions (RQs):
\begin{itemize}
    \item \textbf{RQ1:} Does \ours outperform current state-of-the-art federated recommendation methods?
    \item \textbf{RQ2:} Does \ours mitigate the information loss and preserve the personalized embedding distributions?
    \item \textbf{RQ3:} Does \ours have better scalability across different model architectures?
\end{itemize}

\subsection{Experimental Settings}

\subsubsection{Datasets}

Our experiments are conducted on five widely used datasets: Ml-100k~\cite{harper2015movielens}, Ml-1m~\cite{harper2015movielens}, Beauty~\cite{he2016ups}, Book-Crossing (BX)\cite{ziegler2005improving} and LastFM~\cite{levy2010music}.
These datasets are split into training and test sets in an 8:2 ratio, with 10\% of the training set used as a validation set.
We divided each dataset into 100 clients for federated learning. Considering that some federated recommendation system frameworks are designed with a separate client for each user, we also considered this scenario in the Ml-100k dataset.
The statistics of these datasets are shown in Table~\ref{tab:dataset}.

\subsubsection{Compared Methods}

We compared various baseline methods in our experiments. 
For the independent recommendation algorithm without aggregation, we select three backbone models: MF~\cite{koren2009matrix}, LightGCN~\cite{he2020lightgcn} and GF-CF~\cite{shen2021powerful}. 
For the federated recommendation algorithm, we select three popular methods: FedMF~\cite{chai2020secure}, FedLightGCN, FedNCF~\cite{perifanis2022federated} and two methods that focus on user personalization: Pfedrec~\cite{zhang2023dual} and GPFedRec~\cite{zhang2024gpfedrec}.

\subsubsection{Evaluation Metrics}

We use F1~\cite{herlocker2004evaluating}, Mean Reciprocal Rank (MRR)~\cite{voorhees1999trec} and Normalized Discounted Cumulative Gain (NDCG)~\cite{jarvelin2002cumulated} as evaluation metrics in our experiments (higher values indicate better accuracy). These are popular metrics in the Top-K recommendation scenario. We set $K=10$ for these metrics.

\subsubsection{Implementation Details}

As we introduced in Section~\ref{sec:scal}, \ours has strong scalability and can be applied to both parameter-based and parameter-free methods. To demonstrate this advantage, we applied our method to a simple recommendation system (MF), a graph-based recommendation system (LightGCN), and a parameter-free recommendation system (GF-CF), presenting all experimental results. We also select MLP as the model for a fair comparison with GPFedrec in RQ2.
Our method involves generating a similarity matrix based on item embeddings. For MF and MLP, the item embeddings can be directly used to calculate item similarity. In the case of LightGCN, there are two types of item embeddings: before and after convolution. To avoid the influence of different client graph structures on aggregation, we use the item embeddings before convolution for the calculation.

The implementation of \ours utilizes the PyTorch library on an NVIDIA Tesla T4 GPU. We maintain the same parameter range as \ours and its corresponding independent recommendation algorithm for a fair comparison. Specific parameter details can be found in the reproduced code.

\begin{table}[t]
\centering
\caption{The comparison for different aggregations in Ml-100k dataset.}
\resizebox{1\columnwidth}{!}
{
\begin{tabular}{l|l|ccc}
\toprule
Backbone             & Method     & F1@10           & MRR@10          & NDCG@10         \\ \hline
\multirow{4}{*}{MLP} & IM         & 0.0112          & 0.0748          & 0.1074          \\
                     & WAA        & 0.0993          & 0.4176          & 0.4806          \\
                     & GGA        & 0.1509          & 0.5005          & 0.5748          \\
                     & CIA (ours) & \textbf{0.2176} & \textbf{0.5887} & \textbf{0.6637} \\ \hline
\multirow{4}{*}{MF}  & IM         & 0.0108          & 0.0811          & 0.1151          \\
                     & WAA        & 0.1105          & 0.4330          & 0.4986          \\
                     & GGA        & 0.1203          & 0.4832          & 0.5445          \\
                     & CIA (ours) & \textbf{0.2246} & \textbf{0.5782} & \textbf{0.6593} \\
\bottomrule
\end{tabular}
}
\label{tab:abla}
\end{table}

\subsection{Method Comparison (RQ1)}

We compare \ours with all baseline methods in Table~\ref{tab:result} and observe the following:

1) Both weighted summation aggregation methods and our collaborative information methods outperform the independent training algorithm. This suggests that, in the selected federated recommendation system dataset, information transfer between clients is crucial for better results.
2) Our method consistently outperforms weighted summation aggregation algorithms with the same backbone across all datasets. This improvement is due to our approach's reliance on collaborative information aggregation, which is more effective than parameter aggregation.
3) Pfedrec and GPfedrec address the issue of insufficient personalization caused by diverse client users in federated recommendation algorithms. However, they only consider aggregation at the individual user level, where each client lacks the user relationship graph. Consequently, graph-structured models like LightGCN and GF-CF do not perform effectively. In contrast, our approach extends federated recommendation scenarios to different user groups or companies, where each client can hold multiple users, enhancing the effectiveness of graph-based methods. Additionally, their methods aggregate parameters rather than collaborative information, resulting in inferior performance compared to ours. For a fair comparison, we conduct experiments for a single-user aggregation scenario in the next section (RQ2).
4) Existing federated learning frameworks cannot be directly applied to parameter-free models like graph signal processing models. However, \ours overcomes this limitation by effectively aggregating collaborative information and leveraging the superior performance of graph signal processing in handling graph structures. This enables it to achieve results comparable to those of parameter-based federated recommendation systems.

\begin{figure}[t]
\centering
\includegraphics[width=0.9\columnwidth]{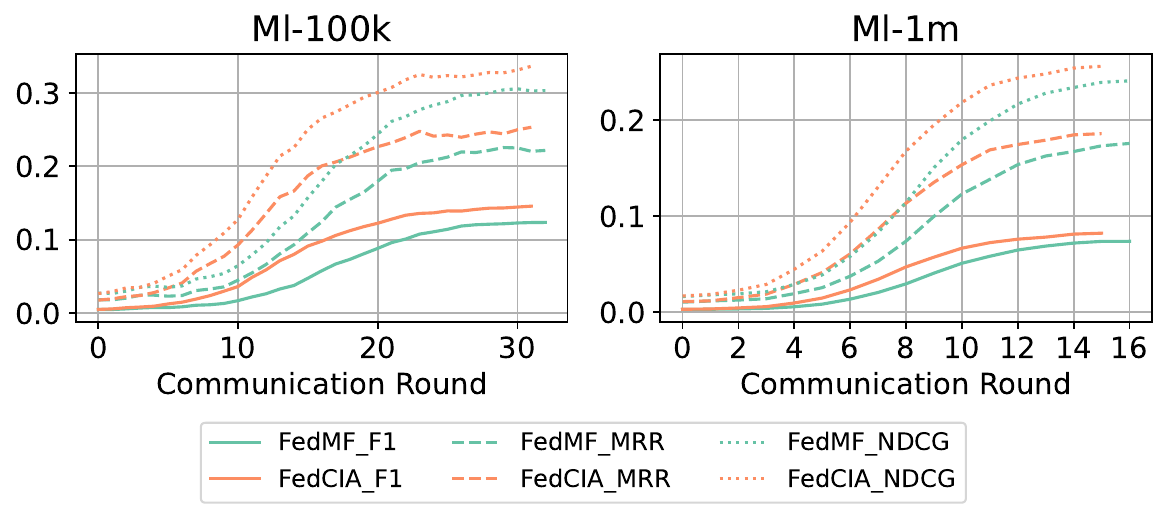}
\caption{
The learning curve for different aggregations on MF in Ml-100k and Ml-1m datasets.
}
\label{fig:con}
\end{figure}

\begin{figure*}[t]
\centering
\includegraphics[width=1.8\columnwidth]{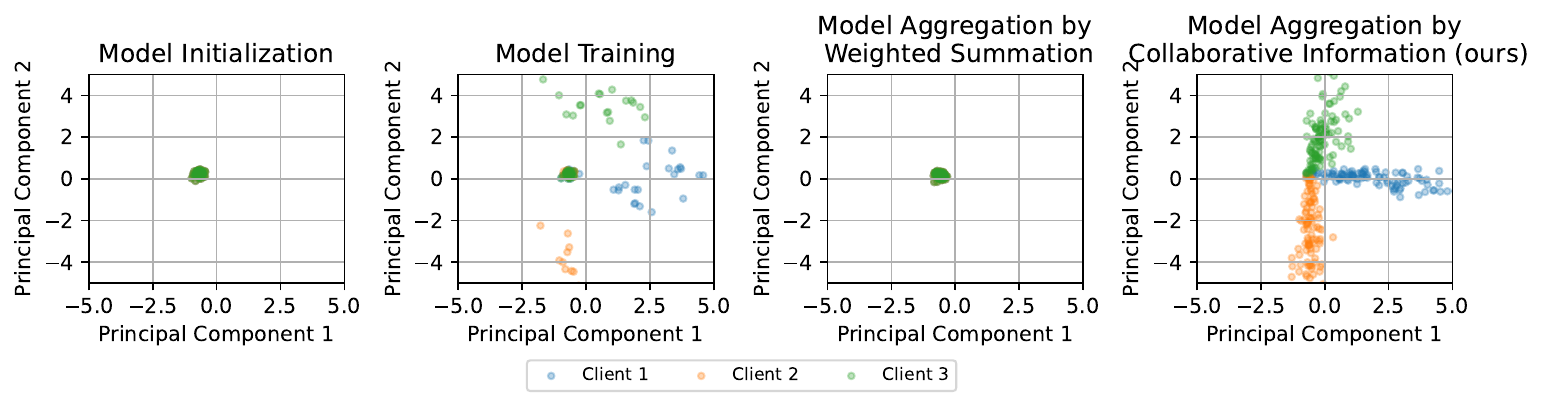}
\caption{
Distribution of item embeddings during different stages of federated learning in different aggregations.
}
\label{fig:aggfig}
\end{figure*}

\begin{figure}[t]
\centering
\includegraphics[width=1\columnwidth]{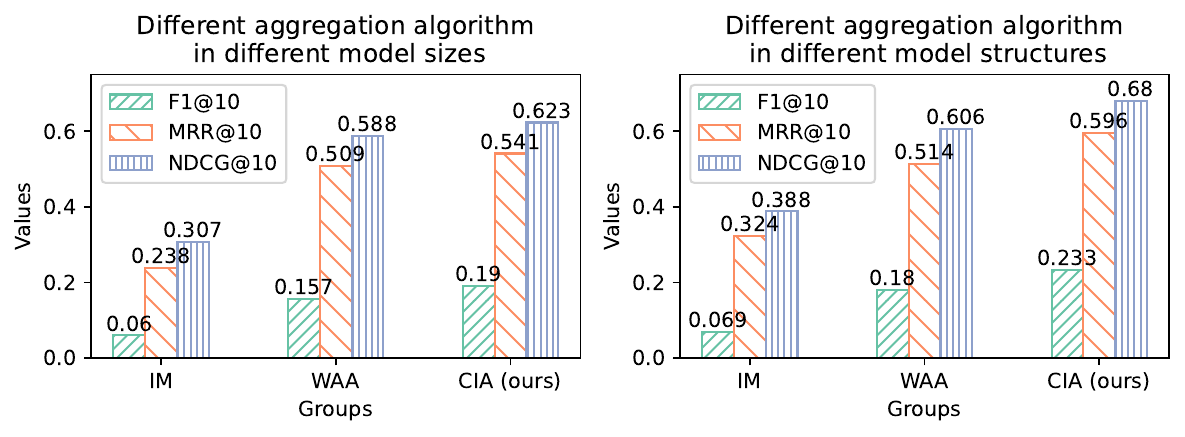}
\caption{
The comparison for different aggregations on heterogeneous scenarios in Ml-100k dataset.
}
\label{fig:he}
\end{figure}

\subsection{Aggregation Comparison (RQ1)}

The core algorithm proposed in \ours is the collaborative information aggregation algorithm, which is an independent module and is challenging to decompose for ablation studies. Instead, we compare our collaborative information aggregation algorithm with other aggregation methods under identical conditions to demonstrate its superiority.
To ensure a fair comparison with personalized federated learning algorithms, we select MLP as the backbone model. Additionally, we configure each client to hold only one user, which accentuates the dispersed data and non-IID challenges, thereby highlighting the benefits of personalized federated learning.
We evaluate the following aggregation algorithms:
\begin{itemize}
\item Independent Model (IM): Each client trains its own model independently without aggregation.
\item Weighted Average Aggregation (WAA): Clients aggregate parameters by weighted summation based on their dataset size (e.g., FedAvg~\cite{mcmahan2017communication}).
\item Graph-Guided Aggregation (GGA): Clients aggregate parameters by forming a relationship graph based on item similarity (e.g., GPFedrec~\cite{zhang2024gpfedrec}).
\item Collaborative Information Aggregation (CIA): Our algorithm.
\end{itemize}

We conducted experiments on the Ml-100k dataset, and the results are presented in Table~\ref{tab:abla}. The results indicate that our collaborative information aggregation consistently outperforms other methods even if we constrain them to the same backbone model. The Independent Model, lacking any aggregation, performs the worst. Weighted Average Aggregation, a classic federated learning approach, improves performance through parameter aggregation. Graph-Guided Aggregation considers personalized client information and models each user individually, achieving notable improvements. However, it still incurs some information loss due to item embedding aggregation. In contrast, our Collaborative Information Aggregation effectively aggregates information while preserving the personalized data from each client.

\subsection{Model Convergence (RQ2)}

In Section~\ref{sec:intro}, we proposed that parameter aggregation leads to more information loss compared to collaborative information aggregation. In this section, we compare the convergence efficiency of them using the Ml-100k and Ml-1m datasets. 
We do not compare graph-guided aggregation for their method used multiple learning rates with significantly different values, making it difficult to make a fair comparison.
The results are illustrated in Figure~\ref{fig:con}.
From the figure, we observe that our approach exhibits better convergence efficiency on both datasets. This demonstrates that the collaborative information aggregation we proposed effectively mitigates the information loss associated with parameter aggregation.

\subsection{Personalized Embeddings (RQ2)}

We conducted a case study on the Ml-100k dataset to evaluate the effectiveness of personalized item modeling, as illustrated in Figure~\ref{fig:agg}. The federated learning algorithm was divided into three stages:
\begin{itemize}
\item Model Initialization: Clients initialize their model with the same distribution.
\item Model Training: Clients train their models using their dataset.
\item Model Aggregation: Clients aggregate their model using a designated aggregation algorithm.
\end{itemize}

We randomly selected three clients and 100 items to examine their respective item distributions across these stages. In the third stage, we compared two aggregation paradigms: the weighted summation aggregation and our proposed collaborative information aggregation. The results are depicted in Figure~\ref{fig:aggfig}.

The figure shows that although each client begins with the same item distribution during initialization, their embeddings diverge significantly after training due to the varied data. The weighted summation aggregation forces all clients into a uniform item distribution, resulting in a loss of personalized information modeling. In contrast, our collaborative information aggregation allows models to integrate collaborative information while maintaining their individual item distributions, effectively leveraging the collaborative data of other items while preserving personalized information, thus achieving better results.

\subsection{Scalability Analysis (RQ3)}

In this section, we conduct experiments on models in different parameter sizes or different model structures to demonstrate the scalability of our proposed aggregation solution.

In our experiment on model size, we redesigned the federated learning scenario, assigning $\frac{1}{3}$ clients to use a MF model with a 128-dimensional embedding, $\frac{1}{3}$ with a 256-dimensional embedding, and $\frac{1}{3}$ with a 512-dimensional embedding. For the experiment on model structure, we selected equal proportions of MF,  MLP, and LightGCN models. Our method is compared with other aggregations, as illustrated in Figure~\ref{fig:he}.

The independent training model, which cannot aggregate any information, performs poorly. The Weighted Average Aggregation can only aggregate clients with identical model architectures, limiting its ability to gather global collaborative information and leading to performance degradation. In contrast, our method effectively utilizes collaborative information from all clients, regardless of differences in model parameter sizes or structures, resulting in superior performance.




%% file: 5-Conclusion.tex
\section{Conclusion}

In this paper, we propose a novel federated aggregation framework, named \ours, for recommendation algorithms. In \ours, we utilize collaborative information, specifically item similarity, to perform information aggregation instead of relying on model parameters. Using graph signal processing theory, we demonstrate that the summation of local item similarities can approximate global collaborative information. Experiments on five popular real-world datasets show that \ours outperforms existing federated recommendation methods in terms of information aggregation, personalization, and scalability.

%% file: Acknowledgments.tex
\begin{acks}

This work is supported by National Natural Science Foundation of China (NSFC) under the Grant No.62172106.
Peng Zhang is a faculty of School of Computer Science, Fudan University.
Tun Lu is a faculty of School of Computer Science, Shanghai Key Laboratory of Data Science, Fudan Institute on Aging, MOE Laboratory for National Development and Intelligent Governance, and Shanghai Institute of Intelligent Electronics \& Systems, Fudan University.

\end{acks}